\documentstyle[12pt, epsfig]{article}  
\begin{document} 
\baselineskip=20pt
\def\la{\mathrel{\mathpalette\fun <}}
\def\ga{\mathrel{\mathpalette\fun >}}
\def\fun#1#2{\lower3.6pt\vbox{\baselineskip0pt\lineskip.9pt
\ialign{$\mathsurround=0pt#1\hfil##\hfil$\crcr#2\crcr\sim\crcr}}}  
\begin{flushright}   
NPI MSU 97-10/46
\end{flushright} 
\vskip 5 cm 
\begin{Large} 
\centerline{\bf RADIATIVE ENERGY LOSS OF HARD PARTON JET} 
\centerline{\bf IN EXPANDING QUARK$-$GLUON FLUID} 
\end{Large} 
\vskip 2 cm 
\centerline{\large \bf I.P.Lokhtin and A.M.Snigirev}  
\vskip 1 cm 
\centerline{\large $Nuclear~Physics~Institute~Moscow~State~University$}
\centerline{\large $119899,~Moscow,~Russia$}   
\vskip 3 cm 
\begin{abstract} 
The gluon radiation induced by multiple scattering of a hard parton in expanding 
quark-gluon plasma is investigated. The radiative energy loss of hard 
parton jet is shown to decrease considerably when angular size of extracted jet goes
up, but the fraction of loss survives even for the reasonable wide cone of angles 
due to the coherent effects taking into account the rescattering of the radiated gluons 
in the QCD medium. The possible application to the jet quenching phenomenon in heavy 
ion collisions at LHC energies is discussed. 
\end{abstract} 
\newpage 
\pagestyle{plain} 
\setcounter{page}{1}

\noindent {\large \bf 1. Introduction}
\bigskip   

One of the most actual goals of high energy physics is the achievement of a
deconfinement of hadronic matter and the study of created quark-gluon plasma (QGP)
properties~\cite{satz}. The jets production and other hard processes are considered as 
effective probes of QGP in future heavy ion collider experiments at RHIC and
LHC~\cite{pqm95}. Since high-$p_T$ parton pair (dijet) from a single hard scattering 
is produced at the very beginning of the collision process ($\la 0.01$ fm/c), 
it propagate through QGP formed due to copious mini-jet production over longer time 
scales ($\sim 0.1$ fm/c) and interacts strongly with the medium constituents. 
The various aspects of hard parton passing through the dense matter
are discussed intensively~\cite{appel}-~\cite{lokhtin}. In particular, the possible 
jet quenching (suppression of a high-$p_T$ jets) in ultrarelativistic
nucleus-nucleus collisions due to significant energy losses of fast jet partons in QGP
is considered~\cite{gyul90,gyul95,baier,lokhtin}.  
      
In our previous paper~\cite{lokhtin} we used to investigate the dijet quenching and
acoplanarity due to the final state interaction (elastic scattering) of high-$p_T$
parton in viscous expanding quark-gluon fluid. So we restricted our discourse to
the collisional energy losses of a hard jet in the medium disregarding radiative one's. 
Although gluon radiation is a very efficient energy loss mechanism for hard parton, 
"bremsstrahlung" gluons are emitted inside the narrow cone of jet $\theta_0 \la 
\sqrt{m_0/E_{jet}}$ in the Bethe-Heitler limit of independent emissions~\cite{ryskin}
($m_0$ is the mean energy of medium constituents);    
and measuring jet energy by the collection of the final hadron energies in the cone 
$\theta \ga \theta_0$ allows the major portion of the energy of initial hard parton to 
be reconstructed in this case. 

The fact is, however, that there is the destructive 
interference between different amplitudes, if gluon radiation induced by multiple 
scattering suppresses soft gluons whose formation time is larger than the mean free 
path of partons scattering in a QCD medium (the Landau-Pomeranchuk effect in analogy to 
QED)~\cite{gyul95,baier}. In particular, it has been shown in paper~\cite{baier} that
the count of diagrams corresponding to scattering of the radiated gluons in the
QCD medium can result in coherent suppression as compared to the Bethe-Heitler
situation ($dE/dx \propto \sqrt{E}$ instead of $dE/dx \propto E$) with gluon emission 
in the wide cone of angles. Anyway, the contribution to the total jet quenching factor
due to radiative energy loss depends on the angular size of the extracted jet cone. 
The aim of our paper is to analyze the angular structure of radiative energy
loss of hard parton jet in expanding quark-gluon fluid and to discuss the possible 
application to the jet quenching phenomenon in heavy ion collisions.  

\newpage 
\noindent {\large \bf 2. The model of jet rescattering in expanding parton matter}
\bigskip   
  
The average number of jet parton scatterings $N_{rsc}$ in a boost-invariant
longitudinal expanding QGP takes the form~\cite{blaizot,lokhtin}  
\begin{equation} 
<N_{rsc}> = \int\limits_0^{2\pi}\frac{d\varphi}{2\pi}\int\limits_0^{R_A}dR\cdot P_A(R) 
\int\limits_{\displaystyle \tau_0}^{\displaystyle \tau_L}d\tau 
\sum_{b}\sigma_{ab}(\tau)\cdot n_b(\tau). 
\end{equation} 
where averaging is performed over jet production vertex ($R$, $\varphi$) and
space-time evolution of the dense matter ($\tau = \sqrt{t^2 - z^2}$); $\tau_0$ and 
$\tau_L$ are the proper times of QGP formation and of jet escape from plasma
respectively, $n_b$ is the number density of plasma constituents of type $b$, 
$\sigma_{ab}$ is the cross section of jet parton $a$ scattering off the constituents 
$b$. The distribution over the distance from the nuclei collision axis $z$ to the 
dijet production vertex R for uniform nucleons density in nucleus $A$ is written
as~\cite{lokhtin}:  
\begin{equation}
P_A(R)  =  \frac{3}{2} \frac{1}{R_A} \left( 1 - \frac{R^2}{R_A^2} \right) , 
~~~R \leq R_A,  
\end{equation} 
where $R_A$ is the colliding nucleus radius. The proper time at which the jet leaves
the plasma $\tau_L > \tau_0$ is calculated as: 
\begin{equation} 
\tau_L = \sqrt{R_A^2-R^2\sin^2{\varphi}} - R\cos{\varphi}.   
\end{equation}
In a linear kinetic theory, the interval between successive rescatterings, $l_i = 
\tau_{i+1} - \tau_i$, is determined according to the probability density:  
\begin{equation} 
\frac{dP}{dl_i} = \lambda^{-1}(\tau_{i+1})\cdot \exp{(-\int\limits_0^{l_i}\lambda^{-1}
(\tau_i + s)ds)},
\end{equation}
where the mean inverse free path $\lambda_a^{-1}(\tau) = \sum_{b}\sigma_{ab}(\tau)
n_b(\tau)$. If the mean free path is much larger than the screening radius in QCD 
medium, $\lambda \gg \mu_D^{-1}$, the successive scatterings are treated as
independent and the effective colour field produced by the plasma constituents 
can be modeled by a static Debye screened potential~\cite{gyul95}. In this case the 
dominant contribution to differential cross section for the scattering of a jet parton 
with energy E off the thermal partons with energy $m_0$ is  
\begin{equation} 
\frac{d\sigma_{ab}}{dt} \cong C_{ab} \frac{2\pi\alpha_s^2(t)}{t^2}, 
\end{equation} 
where $C_{ab} = 9/4, 1, 4/9$ respectively for $gg$, $gq$ and $qq$ scatterings, 
$t$ is the transfer momentum squared,    
$\alpha_s$ is the QCD running coupling constant: 
\begin{equation} 
\alpha_s = \frac{12\pi}{(33-2N_f)\ln{(t/\Lambda^2)}} 
\end{equation} 
for $N_f$ active quark flavours, and $\Lambda \simeq T_c$ is 
the QCD scale parameter ($T_c$ is the critical temperature). The integral parton
scattering cross section 
\begin{equation} 
\sigma_{ab} = \int\limits_{\displaystyle
\mu^2_D(\tau)}^{\displaystyle m_0(\tau)E / 2 }dt\frac{d\sigma_{ab}}{dt} 
\end{equation} 
is regularized by the Debye screening mass evaluated in lowest order 
pQCD~\cite{nadk} $\mu_D^2 \cong 4 \pi \alpha_s^{\ast 2} T^2 (1 + N_f / 6)$ with 
$\alpha_s^{\ast} = \alpha_s (16T^2)$, that is closed to the lattice result of a pure
gauge QCD $\mu_D \simeq 2 T$~\cite{gao} at high temperature limit $T \gg T_c$. 

The average energy loss of the jet parton can be written as sum of contributions from
collisional and radiative loss similar to Eq.(1):      
\begin{eqnarray}
\Delta E_{tot} & = &
\int\limits_0^{2\pi}\frac{d\varphi}{2\pi}\int\limits_0^{R_A}dR\cdot P_A(R)
\cdot \Delta E(R, \varphi), \\    
\Delta E(R, \varphi) & = & \int\limits_{\displaystyle\tau_0}^{\displaystyle 
\tau_L}d\tau \left( \frac{dE}{dx}^{rad}(\tau) + \sum_{b}\sigma_{ab}(\tau)\cdot
n_b(\tau)\cdot \nu(\tau) \right) ,   
\end{eqnarray} 
where the thermal average collisional energy loss of jet parton due to an elastic single
scattering at $E \gg m_0 \sim 3T$ takes the form
\begin{equation}  
\nu(\tau) = <\frac{t}{2m_0}> \simeq \frac{1}{4T(\tau)\sigma_{ab}(\tau)}
\int\limits_{\displaystyle\mu^2_D(\tau)}^
{\displaystyle 3T(\tau)E / 2}dt\frac{d\sigma_{ab}}{dt}t.  
\end{equation} 
Let us mark that the analogous expressions could be written for any average jet
characteristics (for instance, as acoplanarity or scattering angle).      

In order to analyze the structure of radiative losses $dE / dx$, we adapt the 
result of paper~\cite{baier} for the radiative gluon energy spectrum in static 
QCD medium obtained in the soft radiation limit: 
\begin{eqnarray} 
\omega\frac{dI}{d\omega d^2 k_T} & = & \frac{\alpha_s}{\pi^2} \left< 
\left| \sum\limits_{i=1}^{N} {\bf J}_i e^{ik_\mu x_i^\mu}\right| 
^2 \right> = \nonumber \\ 
&  & \frac{\alpha_s}{\pi^2}
\left< 2 Re \sum\limits_{i=1}^{N} \sum\limits_{j=i+1}^{N} 
{\bf J}_i {\bf J}_j \left[ e^{ik_\mu (x_i-x_j)^\mu} -1 \right] 
\>+\> \left| {\sum\limits_{i=1}^{N} {\bf J}_i}\right| ^2\right>, 
\end{eqnarray} 
where {\bf J} is the emission current,  $<>$ is the average over momentum transfers and
longitudinal coordinates of the $N$ scattering centers. The result for $dE / dx$ can be expressed 
formally following to~\cite{baier} as:  
\begin{eqnarray} 
\frac{dE}{dx}^{rad} (\Delta \theta_{jet}) & = & \int\limits_0^Ed\omega
\int\limits_{\displaystyle
\Delta\theta_{jet}^2\omega^2}^{\displaystyle\omega^2}dk_T^2\frac{9\alpha_s}{\pi\lambda}
\frac{\mu_D^2}{k_T^2 (k_T^2 + \mu_D^2)}\cdot \Phi\left[ \frac{\lambda_g k_T^4}{\omega
\mu_D^2}\right] \cdot \Theta\left( \tau_L - \frac{2\omega}{k_T^2}\right) , \\   
\Phi (x) & = & \int\limits_0^{\infty}\frac{dt}{(t+1)^{3/2}} \sin{\frac{xt}{4}}^2.    
\end{eqnarray} 
The integral (12) is carried out over all possible energies $\omega$ and some
effective transverse momentum $k_T$, the direction of which in the general case do not 
converged to the direction of transverse momentum of radiated gluon (with respect to 
the direction of the fast colour charge) due to coherent effects in QCD 
medium~\cite{baier}. The $\Phi$-function appearance is the result of the destructive 
interference between different amplitudes, the $\Theta$-function takes into account the 
finite volume of medium resulting in the formation time of a radiated gluon 
$\tau_f = 2\omega / k_T^2$ can not exceed the time $\tau_L$ at which the jet leaves the 
medium. We regard the radiative loss of jet as "irretrievable" for gluons emitted 
outside some effective cone with $k_T / \omega > \Delta \theta_{jet}$ and as 
"reconstructable" for gluons emitted inside this cone with $k_T / \omega \la \Delta 
\theta_{jet}$. Let us mark that the procedure of introduction of the parameter 
$\Delta \theta_{jet}$ as the effective "wide-angular" emission cone is, of course, 
somewhat artificial (the direction of the colour charge inside the QCD medium is not 
precisely specified~\cite{baier}), and it is done for the search of developing an 
approximate method of qualitative investigation of the jet radiative losses 
structure~\footnote{In the general case the parameter $\Delta \theta_{jet}$ is not 
equal directly to the angular size $\Delta\theta^{det}_{jet} \le \Delta \theta_{jet}$ 
of jet observed in a real experiment and corresponded to the interference of diagrams 
with gluon emission at large and small angles (alought it must be certain relation). In 
fact, the congruence of $\Delta \theta_{jet}$ and $\Delta\theta^{det}_{jet}$ could 
gives one the upper limit of "irretrievable" radiative energy loss for gluons emitted 
outside this cone.}.

We should also remark that we have used the integral representation (13) for function
$\Phi (x)$, although there is more recent derivation~\cite{mueller} resulting to the
Schroedinger-like equation whose "potential" is given by the single-scattering cross
section of hard parton in the medium. The results of both procedures calculating the
radiative energy loss are close (up to logarithmic prefactor in $dE / dx$ in
asymptotic, that is not relevant for our discourse). The possible existence of three 
asymptotic regimes of radiative energy loss in the $E \rightarrow \infty$ limit were 
pointed out in~\cite{baier}: the Bethe-Heitler regime of independent emissions (at 
$\tau_f \ll \lambda$ with $dE / dx \propto E$), the Landau-Pomeranchuk regime of 
coherent suppression (at $\lambda \ll \tau_f \ll \tau_L$ with $dE / dx \propto 
\sqrt{E}$), and the factorization limit (at $\tau_f \gg \tau_L$ with $dE/dx \propto 
const$ up to $\log{E}$ factor). Whether radiation follows the one of the asymptotic 
regimes, depends on $\omega$ and properties of the medium. 

\vskip 1cm   
\noindent {\large \bf 3. A structure of radiative energy loss of parton jet and jet 
quenching in heavy ion collisions}
\bigskip 

We have performed the numerical integration of Eq.(12) for various values of the
jet cone parameter $\Delta \theta_{jet}$ (without any suggestions on asymptotic
behaviour of radiative loss) and generalized the result in accordance with
Eqs.(1)-(10) to the more realistic case realized in symmetric ultrarelativistic nuclei
collisions when created medium is treated as a boost-invariant longitudinal expanding
quark-gluon fluid. The critical point in doing so (which gives the approximate way of 
solution when the time scale of expansion is longer than one of scattering process) is 
to consider both initial and final state gluon radiation associated with each 
scattering in expanding medium together including the interference effect by the 
modified radiation spectrum (12).     

In order to simplify calculations (and not to introduce new parameters) we shall omit 
here the transverse expansion and viscosity of fluid using the well-known scaling 
Bjorken's solution~\cite{bjorken} for temperature and number density of QGP at $T > T_c 
\simeq 200$ MeV: $T(\tau) \tau^{1/3} = T_0 \tau_0^{1/3},~~ n(\tau) \tau = n_0 \tau_0$. 
Let us mark that the transverse flow effect can play an important role in the
formation of final hadronic state at the more later stages of evolution of long-lived
system created in ultrarelativistic nuclei collision; but the influence of transverse 
expansion of QGP, as well as mixed phase at $T = T_c$, on the intensity of jet 
rescattering (that is a strongly increasing function of temperature) seems to be 
inessential for high initial temperatures $T_0 \gg T_c$~\cite{lokhtin}. On the 
contrary, the viscosity presence slows down the cooling rate, which leads to a jet 
parton spending more time in the hottest regions of the medium, and the rescattering 
intensity goes up, i.e., in fact some effective temperature of medium gets shift 
compared to the perfect QGP case~\cite{lokhtin}. Also for certainty we have used the 
initial conditions for gluon dominated plasma formation in central $AA$ collisions at 
LHC energies, that have been estimated perturbatively 
in~\cite{eskola} using the new HERA parton distributions: $\tau_0 \simeq 0.1$ fm/c, 
$T_0 \simeq 1 GeV\cdot (A/207)^{1/6}$, $N_f \simeq 0$. Of course, these estimations  
are rather approximate and model-depending. In particular, the discount of higher 
order $\alpha_s$ terms computing the initial energy density of mini-jet system, 
uncertainties of structure functions in low-$x$ region, and nuclear shadowing effect 
can result in the essential variations of initial energy density of QGP~\cite{eskola}. 

Figure 1 represents the average radiative energy loss of gluon jet ($E_{jet} = 100$, 
$200$ and $300$ GeV) in mid-rapidity region $y=0$ as a function of the parameter 
$\Delta \theta_{jet}$ of jet cone for two different initial temperatures of QGP 
$T_0 = 0.6$ GeV and  $T_0 = 1$ GeV, $A = 207$, $R_A = 1.15A^{1/3}$ fm. We can see that 
the radiative loss at small size of jet cone (parameter $\Delta \theta_{jet}\la 1^0$) 
depends on   
\begin{figure} 
\begin{center} 
\makebox{\epsfig{file=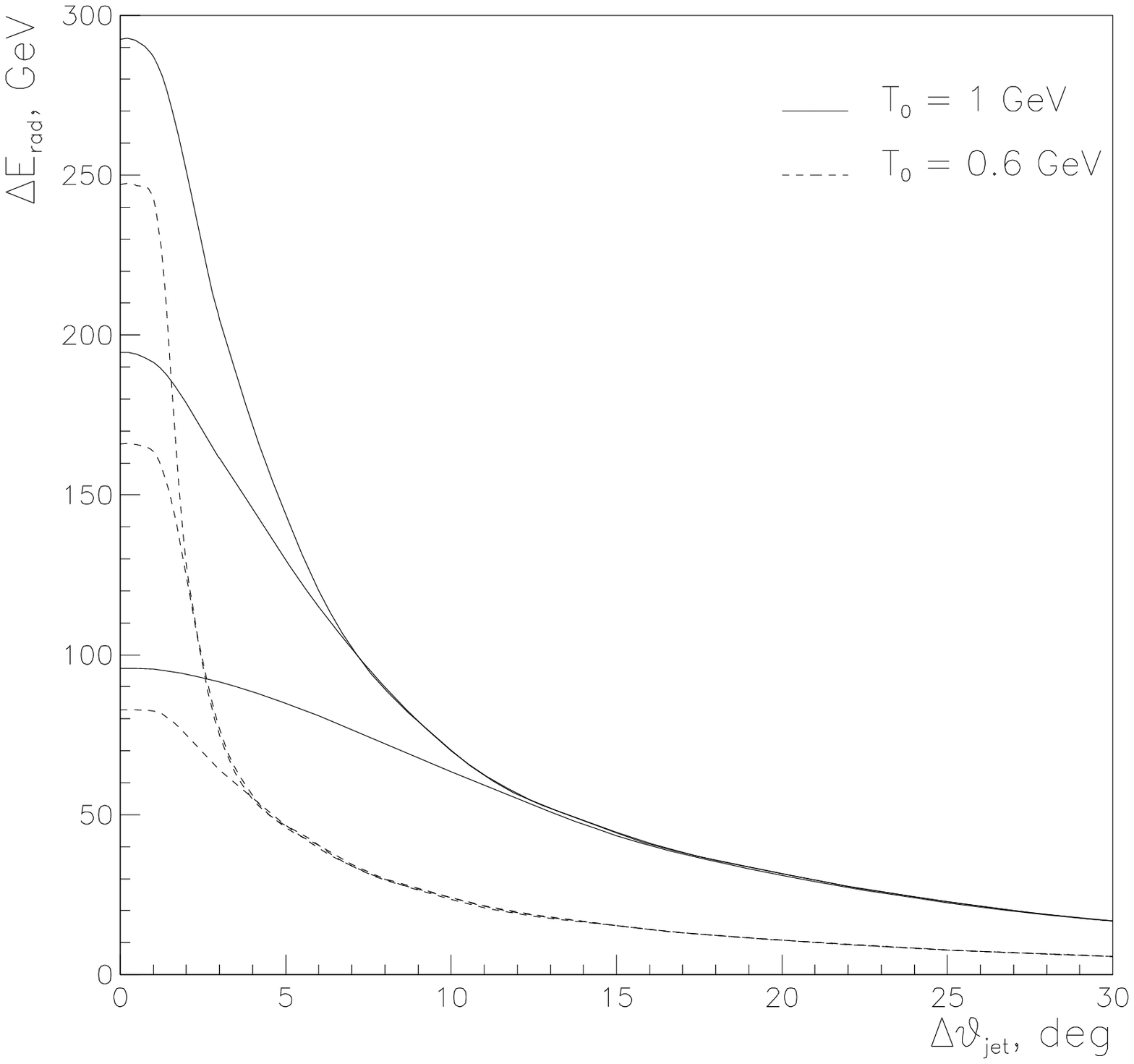, height=100mm}}   
\end{center}
{Figure 1. The average radiative energy loss of gluon jet $\Delta E_{rad}$ in
mid-rapidity region ($y = 0$) as a function of the parameter 
$\Delta \theta_{jet}$ of jet cone, $A = 207$. The upper curves correspond to 
larger initial jet energies ($E_{jet} = 300$, $200$, and $100$ GeV).}        
\end{figure}
\noindent 
medium properties weakly, being determined mainly by the initial
energy of a jet, and it can be significant as compared to the collisional loss: $\Delta
E_{rad} (\Delta \theta_{jet} \rightarrow 0) \sim E_{jet} \gg \Delta E_{col}$ for 
$T_0 = 0.6 - 1$ GeV. The radiative loss of jet decreases rapidly when angular size of 
jet goes up, but the fraction of radiative loss corresponding to regime of coherent 
suppression survives even for the reasonable wide cone of angles (parameter 
$\Delta \theta_{jet} \sim 10^0 - 30^0$), does not almost depend on initial jet energy 
and is determined by the medium properties, that is similar to case of
the collisional loss.   

Finally, we would like to discuss the application of our discourse to a jet quenching 
effect due to the collisional and "wide-angular" radiative energy loss in heavy ion 
collisions at LHC energies. Let us remind that the rate of $\{ij\}$ type dijets with 
$y_1 = y_2 = 0$ and transverse momenta $p_{T1}, p_{T2}$ produced in initial hard 
scattering processes in central $AA$ collisions is 
\begin{eqnarray}
\frac{dN_{ij}^{dijet}}{dy_1dy_2dp_{T1}dp_{T2}} & = & T_{AA}(0)\int\limits_0^{2\pi}
\frac{d\varphi}{2\pi}\int\limits_0^{R_A}dR\cdot P_A(R)\int dp_T^2\frac{d\sigma_{ij}}
{dp_T^2}~\delta(p_{T1} - p_T +  \nonumber \\
 &   & \Delta E_i(\varphi, R))~\delta(p_{T2} - p_T + \Delta E_j(\pi - \varphi, 
R)),
\end{eqnarray} 
where $T_{AA}(b)$ is the nuclear overlap function at zero impact parameter $b = 0$. 
The differential cross section of an initial hard parton-parton scattering in the  
nucleon-nucleon collision can be written as: 
\begin{equation}
\label{hardsec}
\frac{d\sigma_{ij}}{dp_T^2} = K \int dx_1 \int dx_2 \int 
d\widehat{t} f_i(x_1, p^2_T) f_j(x_2, p^2_T) \frac{d\widehat{\sigma}_{ij}}
{d\widehat{t}} \delta (p^2_T - \frac{\widehat{t} \widehat{u}}
{\widehat{s}}) ,   
\end{equation}  
where $s$, $t$ and $u$ are the Mandelstam variables, $f_{i,j}$ are the structure
functions, $x$ is the fraction of momentum which parton carries, the correction factor
$K \sim 2$ takes into account higher order contributions. The 90 degrees CM hard parton 
differential cross section ($y_1 = y_2 = 0$) is calculated using~\cite{gyul90}    
\begin{equation}
\frac{d\sigma_{ij}}{dp_T^2} = \frac{2\pi\alpha_s^2(p_T^2)}{p_T^4}C_{ij}(xf)_
i(xf)_jK,
\end{equation} 
where $(xf)_i \equiv x_T f_i(x_T, p_T^2)$ with $x_T = 2 p_T/\sqrt{s}$, 
the factors $C_{ij}$ include both colour and 90 degrees kinematical factors. 
In order to estimate a jet quenching the $p_T$-dependence of the initial hard parton 
differential cross section was taken as $d\sigma/dp^2_T \propto 
\alpha_s^2(p^2_T)/p^4_T$ for simplicity. The normalized rate of dijets with 
$p_{T1}, p_{T2} > p_{cut}$ in central $AA$ collisions is determined as  
\begin{eqnarray}  
R_{AA}(p_{T1}, p_{T2} > p_{cut}, y = 0) = \frac{\int dy_1\int dy_2\int_{p_{cut}}
dp_{T1}\int_{p_{cut}}dp_{T2}\sum_{i,j}(\frac{\displaystyle dN_{ij}^{dijet}}
{\displaystyle dy_1dy_2dp_{T1}dp_{T2}})_{AA}}{T_{AA}(0)\int dy_1\int dy_2\int_
{{\displaystyle p^2_{cut}}}dp_T^2\sum_{i,j}\frac{\displaystyle d\sigma_{ij}}
{\displaystyle dp_T^2}},   
\end{eqnarray}  
and the jet quenching factor can be introduced as the ratio $R_{AA}/R_{pp} \leq 1$. 

It is worthwhile to notice that the experimental determination of the dijet rates
normalization can be done by the introduction of the reference "unquenched" process, 
for example, as Drell-Yan production ($q \overline{q} \rightarrow \mu^+
\mu^-$), i.e. 
\begin{equation} 
R_{AA} / R_{pp} = \left( \sigma_{AA}^{dijet} / 
\sigma_{pp}^{dijet} \right) /  \left(\sigma_{AA}^{DY} / \sigma_{pp}^{DY} \right). 
\end{equation} 

Figure 2 illustrates the $A$-dependence (Fig.2a, for $T_0 = 1 GeV\cdot (A/207)^{1/6}$) and
$T_0$-dependence (Fig.2b, for $A = 207$) of normalized gluon dijets rate in
mid-rapidity region ($y_1 = y_2 = 0$) when:     
\begin{figure} 
\begin{center} 
\makebox{\epsfig{file=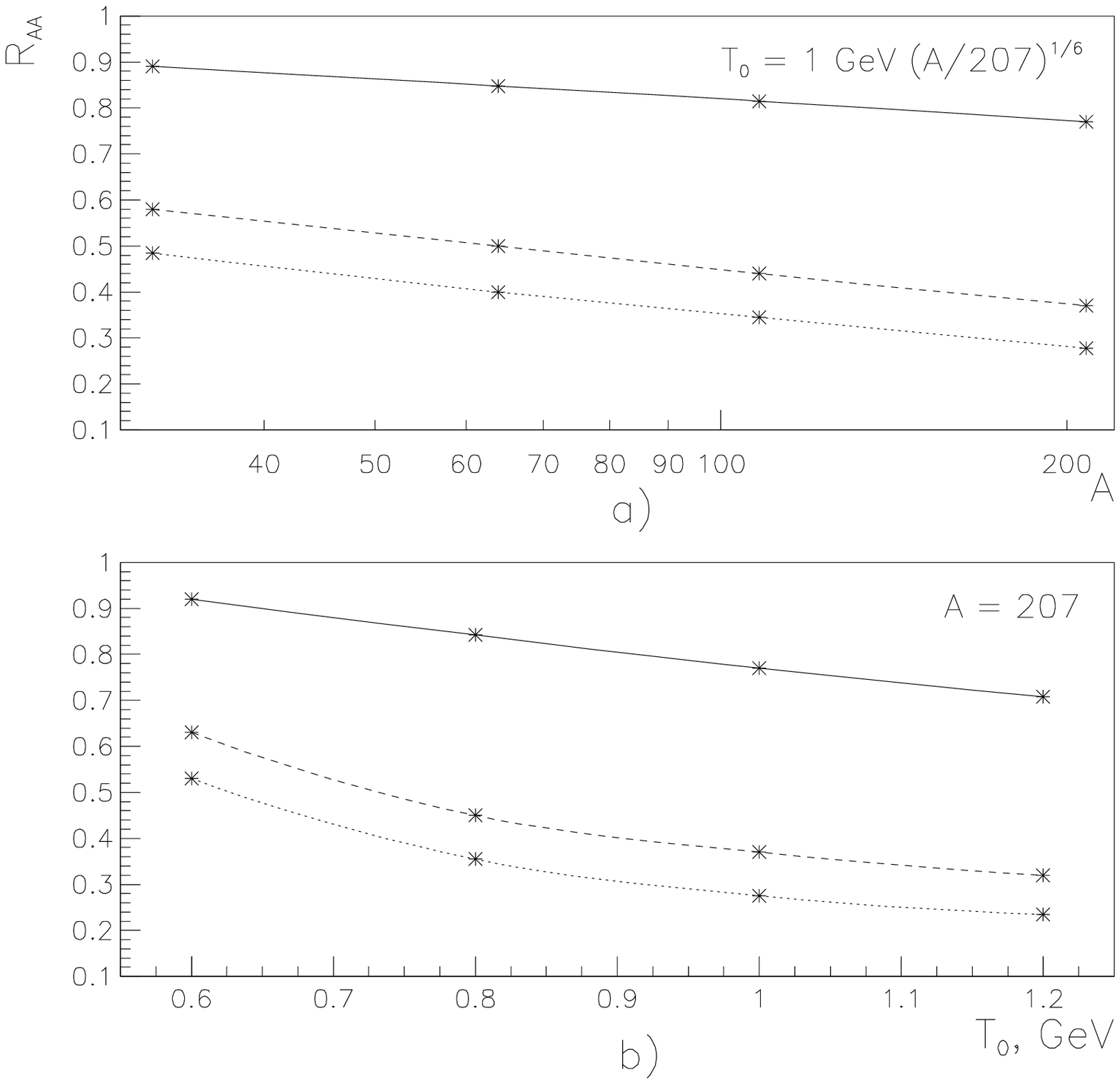, height=100mm}}   
\end{center}      
{Figure 2. The $A$-dependence ($a$) and $T_0$-dependence ($b$) of normalized gluon 
dijets rate in mid-rapidity region ($y_1 = y_2 = 0$) with  $E^{1,2}_{jet} > 100$ GeV. 
Solid curves correspond to the case when just the collisional energy losses was taken 
into account, dashed and dotted curves -- when the collisional and "wide-angular" 
radiative energy losses was taken into account at the values of the parameter 
$\Delta \theta_{jet} = 15^0$ and $\Delta \theta_{jet} = 10^0$ respectively.}  
\end{figure} 
1) the collisional energy losses just was taken into account, \\ 
2) the collisional and "wide-angular" radiative energy losses was taken into account
at jet cone parameter values $\Delta \theta_{jet} = 15^0$ and 
$\Delta \theta_{jet} = 10^0$, \\  
$p_{cut} = 100$ GeV. The significant suppression of hard jets even with finite cone
size due to collisional and "wide-angular" radiative energy losses can be expected at the 
high initial temperatures predicted to be achieved in central heavy ion collisions at 
LHC energies.      

\newpage    
\noindent {\large \bf 4. Conclusions} 
\bigskip 

In summary, we have considered the gluon radiation induced by multiple scattering of a 
hard parton in expanding quark-gluon plasma, taking into account the Landau-Pomeranchuk 
effect in QCD, as well as the finite volume of medium. The radiative energy loss of hard
parton jet with finite cone size is shown to decrease considerably when angular size of 
extracted jet goes up, but the fraction of radiative loss corresponding to scattering 
of the radiated gluons in the medium survives even for the wide cone of angles; it does 
not depend on initial jet energy and is determined by the medium properties. The 
rescattering of jet partons in expanding hot quark-gluon fluid is estimated to reduce 
significantly the dijet rate in mid-rapidity region with $E^{1,2}_{jet} \ga 100$ GeV 
in central heavy ion collisions at LHC energies.

Discussions with R.Baier, V.V.Goloviznin, V.L.Korotkikh, L.I.Sarycheva and G.M.Zinovjev 
are gratefully acknowledged. This research was supported in part by International Soros
Science Education Programme, Grant No A96-194.

\end{document}